\documentclass[twocolumn]{jpsj3}

\usepackage{graphicx}
\usepackage{amssymb,amsmath}

\def\runauthor{Y. Hasegawa and K. Kishigi}

\begin{document}

\title{
Edge states in the three-quarter filled system,
$\alpha$-(BEDT-TTF)$_2$I$_3$
}
%
\author
{Yasumasa Hasegawa and Keita Kishigi$^1$}
\inst
{Department of Material Science, Graduate School of Material Science, 
University of Hyogo, Hyogo, 678-1297, Japan \\
$^1$Faculty of Education, Kumamoto University, Kurokami 2-40-1, 
Kumamoto, 860-8555, Japan}
\date {August 12, 2010}

\abst{
We study the edge states in the two-dimensional 
conductor $\alpha$-(BEDT-TTF)$_2$I$_3$
theoretically. We show that
the Dirac points and the edge states
appear at the $3/4$ and $1/4$ filling 
 as well as the half filling,
due to four sites in the unit cell.
This situation is in contract with the graphene, where
the Dirac points and the edge states appear only at the half filling case.
The edge states exist in both vertical and horizontal edges. 
For the 3/4 filled case 
it is shown that there exists edge states for all possible edges
in the regions of $|k_y| > |K_y|$ or $|k_y| < |K_y|$ for the vertical edges,
and $|k_x| > |K_x|$ or $|k_x| < |K_x|$ for the
horizontal edges, where $\pm(K_x, K_y)$ 
are the positions of the Dirac points in the bulk system,
depending on the choice of edges.
}

\kword{edge state, two-dimension, Dirac point, 
organic conductors, $\alpha$-(BEDT-TTF)$_2$I$_3$, 3/4 filling}
\maketitle
\section{Introduction}
\begin{figure}[b]
\vspace{0.4cm}
\begin{center}
\includegraphics[width=0.45\textwidth]{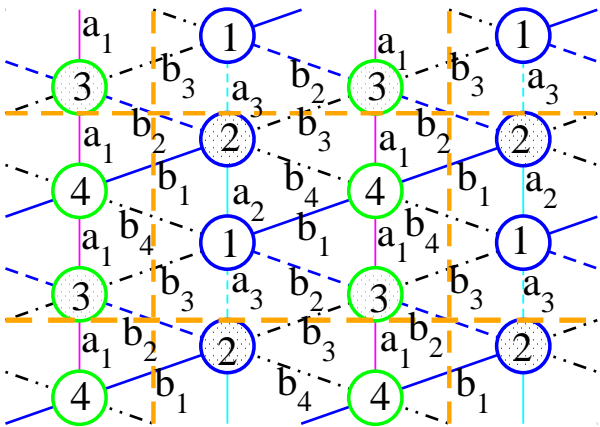}
\end{center}
\caption{(Color online)
The schematic figure of two-dimensional lattice 
for $\alpha$-(BEDT-TTF)$_2$I$_3$.
There are four sites ($1$, $2$, $3$,and $4$) in the unit cell,
three kinds of intra chain transfers ($a_1$, $a_2$ and $a_3$), and 
four kinds of 
inter-chain transfer integrals ($b_1$, $b_2$, $b_3$ and $b_4$).
}
\label{figlattice}
\end{figure}
The massless Dirac particles in the quasi-two-dimensional
organic conductor $\alpha$-(BEDT-TTF)$_2$I$_3$ under 
pressure\cite{Tajima2006}
have been predicted theoretically 
by Katayama, Kobayashi and Suzumura\cite{%
Katayama2006b
}, 
and it has been confirmed by the
interlayer magnetoresistance\cite{Osada2008,Tajima2009}.

The energy dispersion of the massless Dirac particles
\begin{equation}
 E \approx \pm v |\mathbf{k} -\mathbf{k}_0|,
\end{equation}
has been realized in graphene\cite{Novoselov2005,Zhang2005}.
Two bands touch at the Dirac points $\pm \mathbf{k}_0$.
The edge states exist in 
graphene and  have been studied 
by many authors\cite{Fujita1996,Sasaki2006,Peres2006,Kohmoto2007}.

Graphene has a honeycomb lattice structure and it has two
sites in the unit cell. There exists one electron per each site, 
i.e. the band is half-filled in graphene.
On the other hand,
$\alpha$-(BEDT-TTF)$_2$I$_3$ has four sites in the unit cell as shown in
Fig.~\ref{figlattice}, where the small transfers between planes
are neglected. 
The band made from the molecular orbits of
BEDT-TTF molecules is 3/4 filled,
 since one electron is moved from two BEDT-TTF molecules to I$_3$
molecule.
The four sites in the unit cell make the
Dirac points appear 
at $1/4$ and $3/4$ filled cases besides the half-filled case.

We study the $3/4$ filled band in $\alpha$-(BEDT-TTF)$_2$I$_3$
by using the tight-binding model in two dimension. 
The bulk properties of $\alpha$-(BEDT-TTF)$_2$I$_3$
can be expressed by the effective 
two-band model\cite{Kobayashi2007,Georbig2008,Morinari2009}, 
which is similar to the model of the graphene with next-nearest 
hoppings\cite{Georbig2008,Kishigi2008,Kishigi2008b}.
However, when we consider
the edge states, the effective two-band model cannot be applied.

In this paper, we study 
the edge states in
$\alpha$-(BEDT-TTF)$_2$I$_3$ at 3/4 filling 
and compare these edge states with 
the edge states in graphene at half-filling.
We show that the edge states always appear 
either in the region of $|k_y| > |K_y|$ or $|k_y| < |K_y|$
for the vertical edges and
in the region of $|k_x| > |K_x|$ or $|k_x| < |K_x|$
for the horizontal edges, depending on the
choice of the possible edges,
where $k_x$ are $k_y$ are the wave number of the eigenstate and 
$\pm(K_x, K_y)$ are the wave number of the two Dirac points in
the bulk system.

\section{Model}
We study the tight-binding model for the quasi-two-dimensional
conductor $\alpha$-(BEDT-TTF)$_2$I$_3$.
There are four sites in the unit cell, which we label as sites $1$, 
$2$, $3$, and $4$, as shown in Fig.~\ref{figlattice}. 
There are  the hoppings along the $y$ direction, $a_1$, $a_2$ and $a_3$,
and 
the inter-chain hoppings $b_1 - b_4$.
The system has the inversion symmetry\cite{Mori1984,Mori2010,Kondo2009}. 
The inversion center is located at  the site $3$, the site $4$,
and the center of the sites $1$ and $2$. With the inversion the 
sites $1$ and $2$ are exchanged each other 
but the sites $3$ and $4$ remain
unchanged.  
The sites $1$, $2$, $3$ and $4$ are also called 
 AI, AII, 
B and C,  or A, A$'$, B, and C, respectively.
We also take account the site energies $\epsilon_1$, $\epsilon_2$,
 $\epsilon_3$ and $\epsilon_4$. When $\epsilon_1=\epsilon_2$, 
the inversion symmetry is conserved.\cite{Kondo2009}
The energy of electrons in the tight binding 
approximation is given by the equation
\begin{align}
 &   b_1 \psi_{n,m}^{(4)}   + b_2 \psi_{n,m}^{(3)}
 +   b_3 \psi_{n-1,m}^{(3)} + b_4 \psi_{n-1,m}^{(4)} \nonumber \\
 &+ a_2 \psi_{n,m}^{(2)}    + a_3 \psi_{n,m-1}^{(2)} 
  +\epsilon_1  \psi_{n,m}^{(1)}
 = E  \psi_{n,m}^{(1)},
\\
 &    b_1 \psi_{n-1,m}^{(4)}  + b_2 \psi_{n-1,m+1}^{(3)}
 +    b_3 \psi_{n,m+1}^{(3)}  + b_4 \psi_{n,m}^{(4)} \nonumber \\
 &+ a_2 \psi_{n,m}^{(1)}      + a_3 \psi_{n,m+1}^{(1)} 
  +\epsilon_2  \psi_{n,m}^{(2)}
 = E  \psi_{n,m}^{(2)},
\\
 &  b_2 \psi_{n,m}^{(1)}    + b_2 \psi_{n+1,m-1}^{(2)}
 +  b_3 \psi_{n+1,m}^{(1)}  + b_3 \psi_{n,m-1}^{(2)} \nonumber \\
 &+ a_1 \psi_{n,m}^{(4)}    + a_1 \psi_{n,m-1}^{(4)} 
  +\epsilon_3  \psi_{n,m}^{(3)}
= E  \psi_{n,m}^{(3)},
\\
 &   b_1 \psi_{n,m}^{(1)}   + b_1 \psi_{n+1,m}^{(2)}
 +   b_4 \psi_{n+1,m}^{(1)} + b_4 \psi_{n,m}^{(2)} \nonumber \\
 &+ a_1 \psi_{n,m}^{(3)}    + a_1 \psi_{n,m+1}^{(3)} 
 +\epsilon_4  \psi_{n,m}^{(4)}
= E  \psi_{n,m}^{(4)},
\end{align}
where $\psi_{n,m}^{(1)} - \psi_{n,m}^{(4)}$ are
the wave functions 
at sites $1 - 4$
in the unit cell $(n,m)$ with integers $n$ and $m$.
The parameters are estimated by Kondo et al.\cite{Kondo2005,Kondo2009}
 and used by Kobayashi et al.\cite{Kobayashi2009} as 
$a_1=-0.028-0.0025 p$, $a_2=0.048+0.008 p$, 
$a_3=-0.020+0.0005 p$, $b_1=0.123$, $b_2=0.140+0.0015 p$,
$b_3=-0.062-0.002 p$ and $b_4=-0.025$, where $p$ is the uniaxial strain
in the $y$ direction.  
In this paper we use the parameter at $p=4$ kbar, i.e. $a_1=-0.038$,
$a_2=0.08$, $a_3=-0.018$, $b_1=0.123$, $b_2=0.146$, $b_3=-0.07$ and
$b_4=-0.025$.

\section{Bulk system}
\begin{figure}[bt]
\begin{center}
\includegraphics[width=0.48\textwidth]{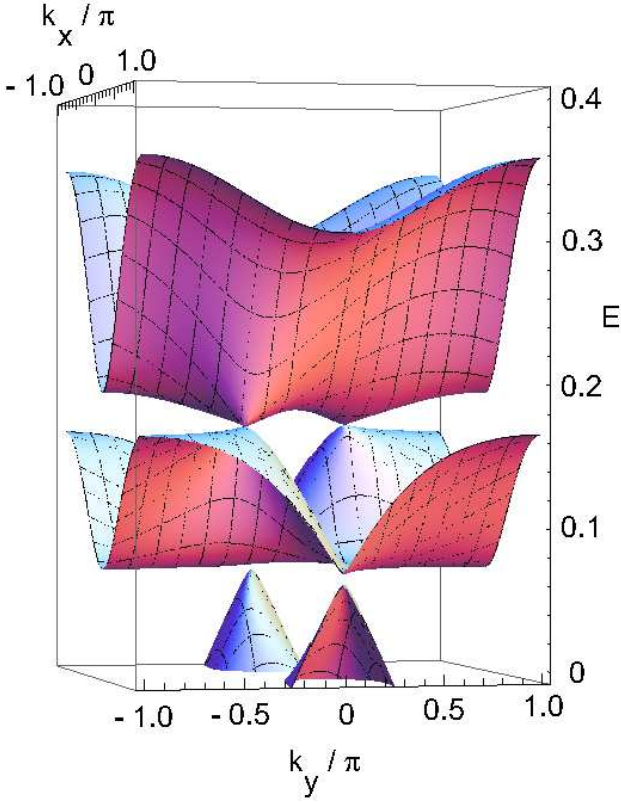}
\end{center}
\caption{(Color online)
The Dirac cone of $\alpha$-(BEDT-TTF)$_2$I$_3$. 
The upper two bands touch at the Dirac points,
$\pm(K_x, K_y)= \pm (0.398 \pi, -0.353 \pi)$}
\label{figfig3d}
\end{figure}
If the system is infinite or periodic with respect to both
$x$ and $y$ directions, the eigenstates are written as
\begin{equation}
\left( \begin{array}{c}
\psi_{n,m}^{(1)} \\ \psi_{n,m}^{(2)} \\
\psi_{n,m}^{(3)} \\ \psi_{n,m}^{(4)}
 \end{array} \right)
= e^{i (k_x n + k_y m)}
\left( \begin{array}{r}
                                         \Phi_{\mathbf{k}}^{(1)} \\
e^{i                   \frac{1}{2} k_y } \Phi_{\mathbf{k}}^{(2)} \\
e^{i (\frac{1}{2}k_x - \frac{1}{4} k_y)} \Phi_{\mathbf{k}}^{(3)} \\
e^{i (\frac{1}{2}k_x + \frac{1}{4} k_y)} \Phi_{\mathbf{k}}^{(4)}
\end{array} \right).
\end{equation}
Then the energy is obtained by 
\begin{equation}
 H_0 \Phi_{\mathbf{k}} = E_0 \Phi_{\mathbf{k}},
\end{equation}
where
\begin{equation}
\Phi_{\mathbf{k}}=
 \left( \begin{array}{c}
\Phi_{\mathbf{k}}^{(1)} \\
\Phi_{\mathbf{k}}^{(2)} \\
\Phi_{\mathbf{k}}^{(3)} \\
\Phi_{\mathbf{k}}^{(4)}
\end{array} \right),
\end{equation}
\begin{equation}
 H_0= \left( \begin{array}{cccc}
\epsilon_1  & C_2         & D_2        & D_1   \\
C_2^*       & \epsilon_2  & D_2^*      & D_1^* \\
D_2^*       & D_2         & \epsilon_3 & C_1   \\
D_1^*       & D_1         & C_1        & \epsilon_4
 \end{array} \right),
\label{eqbulkH}
\end{equation}
\begin{align}
C_1 &= 2 a_1 \cos \frac{k_y}{2}, \\
C_2 &= a_2 e^{i\frac{k_y}{2}} + a_3 e^{-i \frac{k_y}{2}}, \\
D_1 &= b_1 e^{i( \frac{k_x}{2} + \frac{k_y}{4})} 
    +  b_4 e^{i(-\frac{k_x}{2} + \frac{k_y}{4})},  \\
D_2 &= b_2 e^{i( \frac{k_x}{2} - \frac{k_y}{4})} 
    +  b_3 e^{i(-\frac{k_x}{2} - \frac{k_y}{4})}.  
\end{align}
If $a_1=a_2=a_3=0$ (i.e. $C_1=C_2=0$)
and $\epsilon_1=\epsilon_2=\epsilon_3=\epsilon_4=0$,
 the eigenvalues of the matrix in
 Eq.~(\ref{eqbulkH}) are obtained by Mori\cite{Mori2010} as
\begin{equation}
 E_{\mathbf{k}} =\pm \sqrt{ |D_1|^2+|D_2|^2 
  \pm \sqrt{ (D_1^2 +  D_2^2) (D_1^{*2}+ D_2^{*2})}}.
\end{equation}
The condition for the Dirac points at the 1/4 and 3/4 filled band
is obtained for this case of $C_1=C_2=0$ as,
\begin{equation}
 D_1 = \pm i D_2.
\end{equation}
This condition is fulfilled at $(k_x, k_y) = (K_x, K_y)$, where
$K_x$ and $K_y$ is given by
\begin{equation}
e^{i\frac{K_y \mp \pi}{2}} = 
\frac{b_2 e^{i K_x} + b_3}{b_1 e^{i K_x} + b_4}.
\label{eqzero}
\end{equation}
In order to have real solutions,
the absolute value of the right hand side
in Eq.~(\ref{eqzero}) should be unity, from which we obtain
\begin{equation}
 \cos K_x = \frac{b_1^2+b_4^2-b_2^2-b_3^2}{2 (b_2 b_3 - b_1 b_4)},
\end{equation}
which has been obtained by Mori\cite{Mori2010}.

When $C_1$ and $C_2$ are not zero, which is the case in
$\alpha$-(BEDT-TTF)$_2$I$_3$,
the eigenvalues are complicated, although the 
analytical expression is possible as a solution of the quartic equation. 

In Fig.~\ref{figfig3d}, we show   the 3D plot of the energy 
with the parameters for $\alpha$-(BEDT-TTF)$_2$I$_3$
with $p=4$ kbar and $\epsilon_1=\epsilon_2=\epsilon_3=\epsilon_4=0$. 
In this figure we show  the fourth and the third bands from the bottom 
of the energy (i.e. the first and the second band from the top
of the energy)
and the part of the second band 
from the bottom of the energy (the third band from the top of the energy).
The first and the second bands
from the top of the energy
 touch at two Dirac points $\pm(K_x, K_y)= \pm (0.398 \pi, -0.353 \pi)$.
In this choice of parameters a finite gap exists between the second
and the third bands as shown in Fig.~\ref{figfig3d}.
In this paper, however, we focus on the edge states in the
three-quarter filled band and do not study the half-filled case,
since we are interested in 
$\alpha$-(BEDT-TTF)$_2$I$_3$, which is 
the system with the 3/4-filled band.

\section{Edge states}
%
\subsection{vertical edges}
In this section we study 
the system with edges.
The canted edges may be possible, 
but we consider the vertical and horizontal edges
in this paper.

First we consider the vertical edges.
We study two possibilities for each 
 edge.
The left edge and the right edge can be either 
the chain of the sites 1 and 2, or the
chain of the sites 3 and 4 (see Fig.~\ref{figlattice}).
Consider the ribbon of $2 \times L$ chains 
(where $L$ is integer)
 in which 
the left edge is the chain of sites $1$ and $2$, 
and the right edge are the chain 
of the sites $3$ and $4$. We call this system as the (12-34) edge.
The other choices of the edges we will study for the vertical edges are
the boundaries with the chain of the sites $3$ and $4$ 
at the left edge and the chain of the sites $1$ and $2$  at the right edge,
which we call the (34-12) edge, 
the boundaries with chains of the sites $1$ and $2$ 
at both edges (the (12-12) edge),
and  the boundaries with chains of the sites $3$ and $4$ and 
at  both edges, (the (34-34) edge). 
Note that the (34-12) edge has $2 L$ chains, while the (12-12) edge
 and the (34-34) edge have $2L +1$ chains.

We assume the periodic boundary conditions in the $y$ direction
for the systems with vertical edges,
which means that we study the very long vertical ribbon or the tube 
similar to carbon nanotube.

Similar to the honeycomb lattice\cite{Kohmoto2007},
we perform the Fourier transformation with respect to 
$y$ as
\begin{equation}
\left( \begin{array}{c}
\psi_{n,m}^{(1)} \\ \psi_{n,m}^{(2)} \\
\psi_{n,m}^{(3)} \\ \psi_{n,m}^{(4)}
 \end{array} \right)
= e^{i  k_y m}
\left( \begin{array}{r}
                        \Psi_{n,k_y}^{(1)} \\
e^{ i \frac{1}{2} k_y } \Psi_{n,k_y}^{(2)} \\
e^{-i \frac{1}{4} k_y } \Psi_{n,k_y}^{(3)} \\
e^{ i \frac{1}{4} k_y } \Psi_{n,k_y}^{(4)}
\end{array} \right).
\end{equation}
The energy and the eigenstate are obtained by the equation
\begin{equation}
 \left( \begin{array}{cccccc}
M        & N      &\mathbf{0}& \cdots      &\mathbf{0}  &\alpha N^* \\
N^{\dagger}& M    & N      & \mathbf{0}  & \cdots     &\mathbf{0} \\
\mathbf{0} & \ddots   & \ddots   &  \ddots     &\mathbf{0}  & \vdots    \\
\vdots     &\mathbf{0}&\ddots    &\ddots       &\ddots      & \mathbf{0}\\
\mathbf{0} & \cdots   &\mathbf{0}& N^{\dagger} & M& N  \\
\alpha N &\mathbf{0}&\mathbf{0}&\mathbf{0}   & N^{\dagger}& M  \\
\end{array}
\right) \Psi = E \Psi,
\label{eqharper}
\end{equation}
where 
$\mathbf{0}$, $M$, $N$,
are $4 \times 4$ matrix given by
\begin{equation}
 \mathbf{0}=\left( \begin{array}{cccc}
 0&0&0&0 \\
 0&0&0&0 \\
 0&0&0&0 \\
 0&0&0&0  
\end{array} \right),
\end{equation}
\begin{equation}
M= \left( \begin{array}{cccc}
 \epsilon_1 & A_{2} & B_{2}^*            & B_{1}\\
 A_{2}^*& \epsilon_2      & B_{3}  & B_{4}^* \\
 B_{2}    & B_{3}^* & \epsilon_3 & A_{1} \\
 B_{1}^*& B_{4}       & A_{1} & \epsilon_4 \\
 \end{array} \right) ,
\label{eq44matrix}
\end{equation}
\begin{equation}
N_n= \left( \begin{array}{cccc}
0 & 0 & 0 & 0 \\
0 & 0 & 0 & 0 \\
 B_{3}&  B_{2}^* & 0 & 0 \\
 B_{4}^*            &  B_{1} & 0 & 0  
 \end{array} \right) ,
\end{equation}
\begin{align}
 A_{1} &= 2 a_1\cos \frac{k_y}{2}   ,\\
 A_{2} &= a_2 e^{ i \frac{k_y}{2} } 
         +a_3 e^{-i \frac{k_y}{2} } , \\
 B_{1} &= b_1 e^{i  \frac{k_y}{4} } , \\
 B_{2} &= b_2 e^{i  \frac{k_y}{4} } ,\\
 B_{3} &= b_3 e^{i  \frac{k_y}{4} } ,\\
 B_{4} &= b_4 e^{i  \frac{k_y}{4} } ,
\end{align}
%
and $\Psi$ is the vector with $4 L$ wave functions,
\begin{equation}
\Psi= 
\left( \begin{array}{c}
   \Psi_{1,k_y}^{(1)} \\ \Psi_{1,k_y}^{(2)} \\
   \Psi_{1,k_y}^{(3)} \\ \Psi_{1,k_y}^{(4)} \\
\vdots \\
   \Psi_{n,k_y}^{(1)} \\ \Psi_{n,k_y}^{(2)} \\ 
   \Psi_{n,k_y}^{(3)} \\ \Psi_{n,k_y}^{(4)} \\
\vdots \\
   \Psi_{L,k_y}^{(1)} \\ \Psi_{L,k_y}^{(2)} \\ 
   \Psi_{L,k_y}^{(3)} \\ \Psi_{L,k_y}^{(4)} 
\end{array} \right) .
\end{equation} 
In Eq.~(\ref{eqharper}), we had introduced the parameter $\alpha$,
which describe the boundary conditions.
For the open boundary condition, in which the edge states may exist,
we should take $\alpha=0$. 
The periodic boundary conditions can be obtained by taking $\alpha=1$.

In the same way we take the boundary with the 3 and 4 sites
at the left edge and 1 and 2 sites at the right edge,
which we call (34-12) edge.
The energy of the (34-12) edge is the same as that of the 
(12-34) edge except that the left and the right edge states are exchanged.

In the (12-12) edge and the (34-34) edge, we cannot apply the periodic
boundary conditions in the $x$ direction.
The matrix size is $(4L+2) \times (4L+2)$ in these edges.

Note that the (12-12) edge and the (34-34) edges are symmetric with respect
to inversion, while 
the (12-34) edge and the (34-12) edge are not symmetric
but they are exchanged each other by inversion.

\subsection{horizontal edges}

For the horizontal edges, 
the lower and upper boundaries have the zigzag shape.
Each edge consists with 
one pair of the four
possibilities, i.e.,
sites 4 and 2, 2 and 3, 3 and 1, or 1 and 4. 
There are 16 possibilities for the choice of the horizontal edges.
We call these 16 possible edges as the (42-42) edge, 
the (42-23) edge, and so on.
Considering that the sites 1 and 2 are exchanged
each other by inversion
 while sites 3 and 4 are not changed by inversion, 
we obtain that 
the (42-14) edge, the (23-31) edge, the (31-23) edge and 
the (14-42) edge are symmetric with respect to inversion.
The (31-42) edge, for example, becomes the (14-23) edge
by inversion. 
For the horizontal edges, we can perform the 
Fourier transformation with respect to $x$
and the energy is labeled by $k_x$.

There are $4L$ eigenstates for each $k_x$ in the
(42-31) edge, the (23-14) edge, the (31-42) edge and the
(14-23) edge, where integer $L$ is the number of each sites 
($1 - 4$) in each vertical chain. In these four cases we can apply the 
periodic boundary conditions with respect to $y$ direction,
as in the cases of the (12-34) edge and the (34-12) edge
for the vertical edges. If the periodic boundary conditions 
with respect to $k_y$ direction are taken, the above four 
cases give the same energy dispersion.

 There are $4L+1$ eigenstates
 for the (42-14) edge, the (23-42) edge, (31-23) edge and 
the (14-31) edge, where the lowermost and the uppermost
sites are the same. When the lowermost 
and the second lower sites are the same as the second 
upper and the uppermost sites, 
such as the (42-42) edge, the (23-23) edge, the (31-31) edge and the 
(14-14) edge, there are $4L+2$ eigenstates for each $k_x$.
When the second lower sites are the same as the second upper sites,
such as
the (42-23) edge, the (23-31) edge, the (31-14) edge and the
(14-42) edge, there are $4L+3$ eigenstates for each $k_x$.

\section{Results}
\begin{figure}[btp]
\begin{center}
\includegraphics[width=0.48\textwidth]{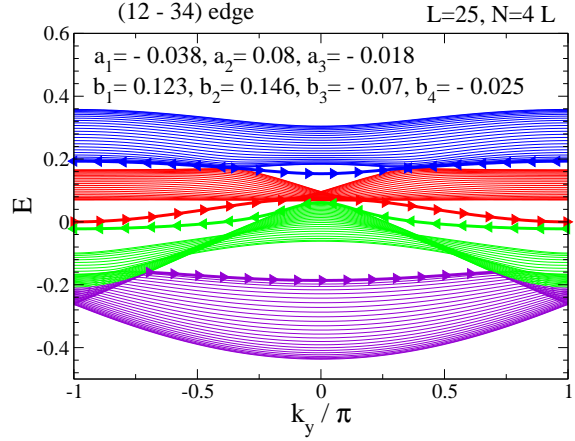}
\end{center}
\caption{(Color online)
The energy spectrum in the system with four sites in
the unit cell with the (12 - 34) edge.
We take the parameters for the $\alpha$-(BEDT-TTF)$_2$I$_3$
under the uniaxial pressure $4$ kbar in the $a$-direction.
There are four bands, each of which contains the $L$ eigenvalues.
The different brightness of lines (purple, green, red, and blue
in the online version) 
represents the different bands.
Thick lines with left  triangles are the edge states on the
left edge, and thick lines with right triangles are the edge states
on the right edge.
}
\label{figedge1234}
\end{figure}
\begin{figure}[btp]
\begin{center}
\begin{flushleft}(a)\end{flushleft} 
\includegraphics[width=0.38\textwidth]{64500fig4a.eps}
\begin{flushleft}(b)\end{flushleft} 
\includegraphics[width=0.38\textwidth]{64500fig4b.eps}
\begin{flushleft}(c)\end{flushleft} 
\includegraphics[width=0.38\textwidth]{64500fig4c.eps}
\begin{flushleft}(d)\end{flushleft} 
\includegraphics[width=0.38\textwidth]{64500fig4d.eps}
\end{center}
\caption{(Color online)
Energy spectrum near $\frac{3}{4}$ filling
for systems with periodic boundary and with edges.
The parameters are the same as in Fig.~\ref{figedge1234}.
}
\label{figedgee0}
\end{figure}
\begin{figure}[btp]
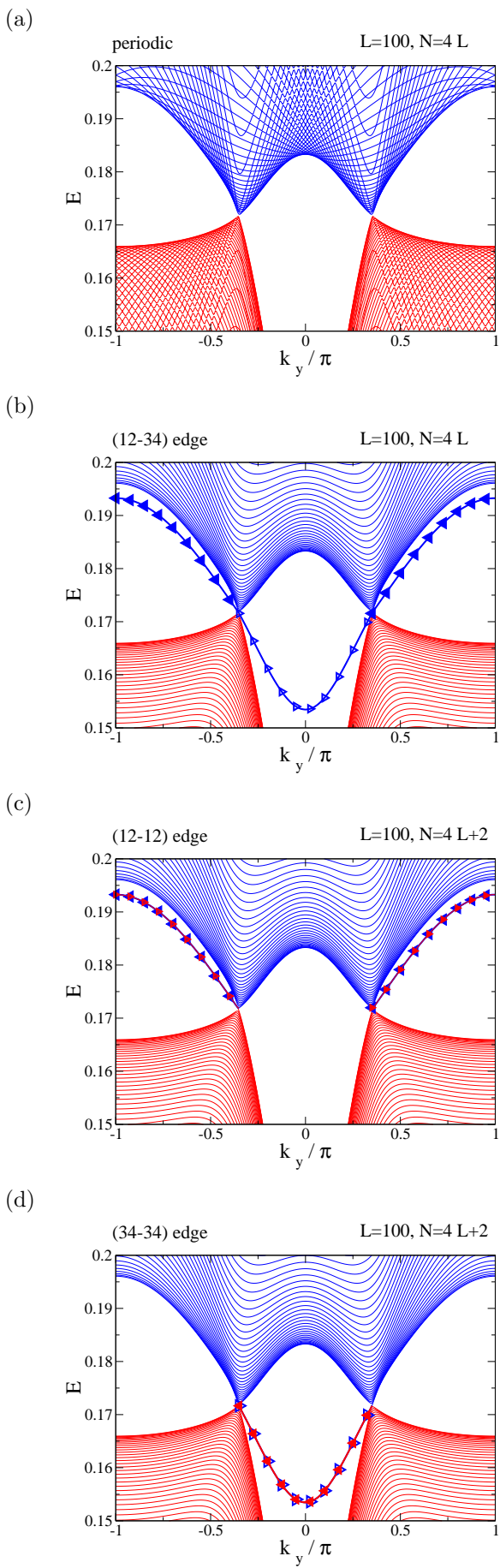
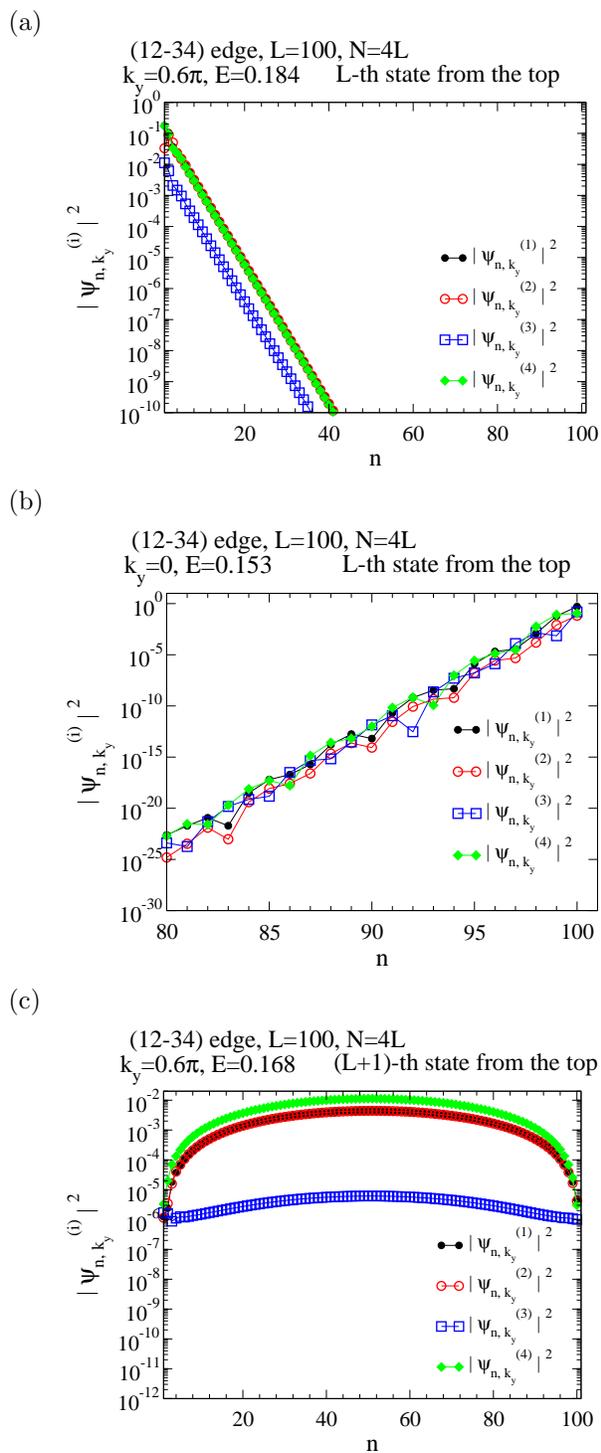

\begin{center}
\begin{flushleft}(a)\end{flushleft}\vspace{-0cm}
\includegraphics[width=0.4\textwidth]{64500fig5a.eps}
\begin{flushleft}(b)\end{flushleft}\vspace{-0cm}
\includegraphics[width=0.4\textwidth]{64500fig5b.eps}
\begin{flushleft}(c)\end{flushleft}\vspace{-0cm}
\includegraphics[width=0.4\textwidth]{64500fig5c.eps}
\end{center}
\caption{(Color online)
Edge states near $\frac{3}{4}$ filling
for systems with $(12-34)$ edges with $L=100$.
The squares of the absolute value
$\Psi_{n,k_y}^{(i)}$ at (a) $k_y=0.6\pi$ and (b) $k_y=0$
of the $L$-th  state from the top, 
and  at (c) $k_y=0.6 \pi$ of $(L+1)$-th  state from the top
are plotted as a function of $n$.
The values of $|\Psi_{n,0.6\pi}^{(1)}|^2$,
$|\Psi_{n,0.6\pi}^{(2)}|^2$ and $|\Psi_{n,0.6\pi}^{(4)}|^2$ in (a)
and the values of  $|\Psi_{n,0.6\pi}^{(1)}|^2$
 and $|\Psi_{n,0.6\pi}^{(2)}|^2$ in (c)
are almost the same in this scale of the figure.
The exponentially localized character is  seen
in (a) and (b).}
\label{figwavef1234}
\end{figure}
\begin{figure}[btp]
\begin{center}
\begin{flushleft}(a)\end{flushleft}
\includegraphics[width=0.4\textwidth]{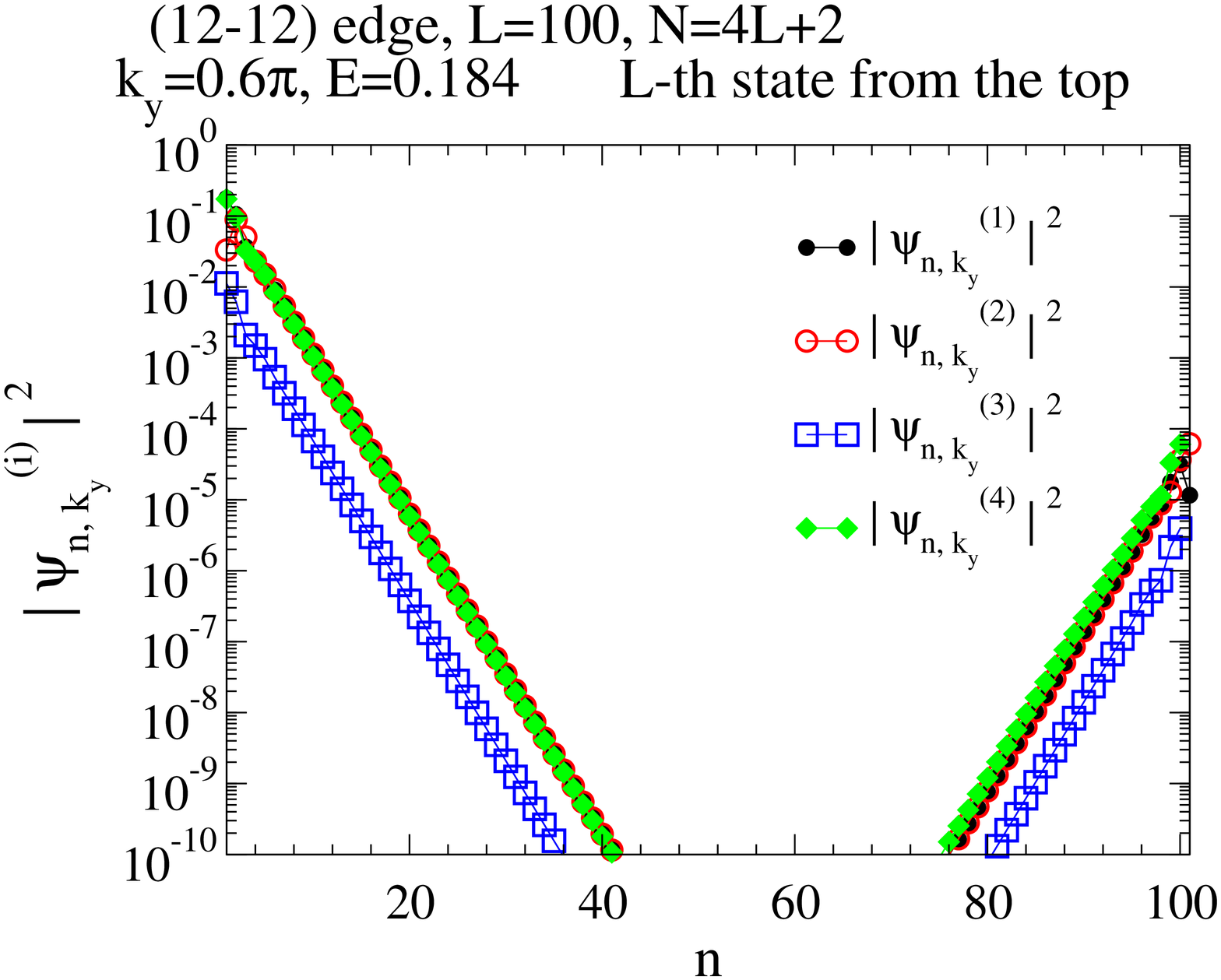}
\begin{flushleft}(b)\end{flushleft}
\includegraphics[width=0.4\textwidth]{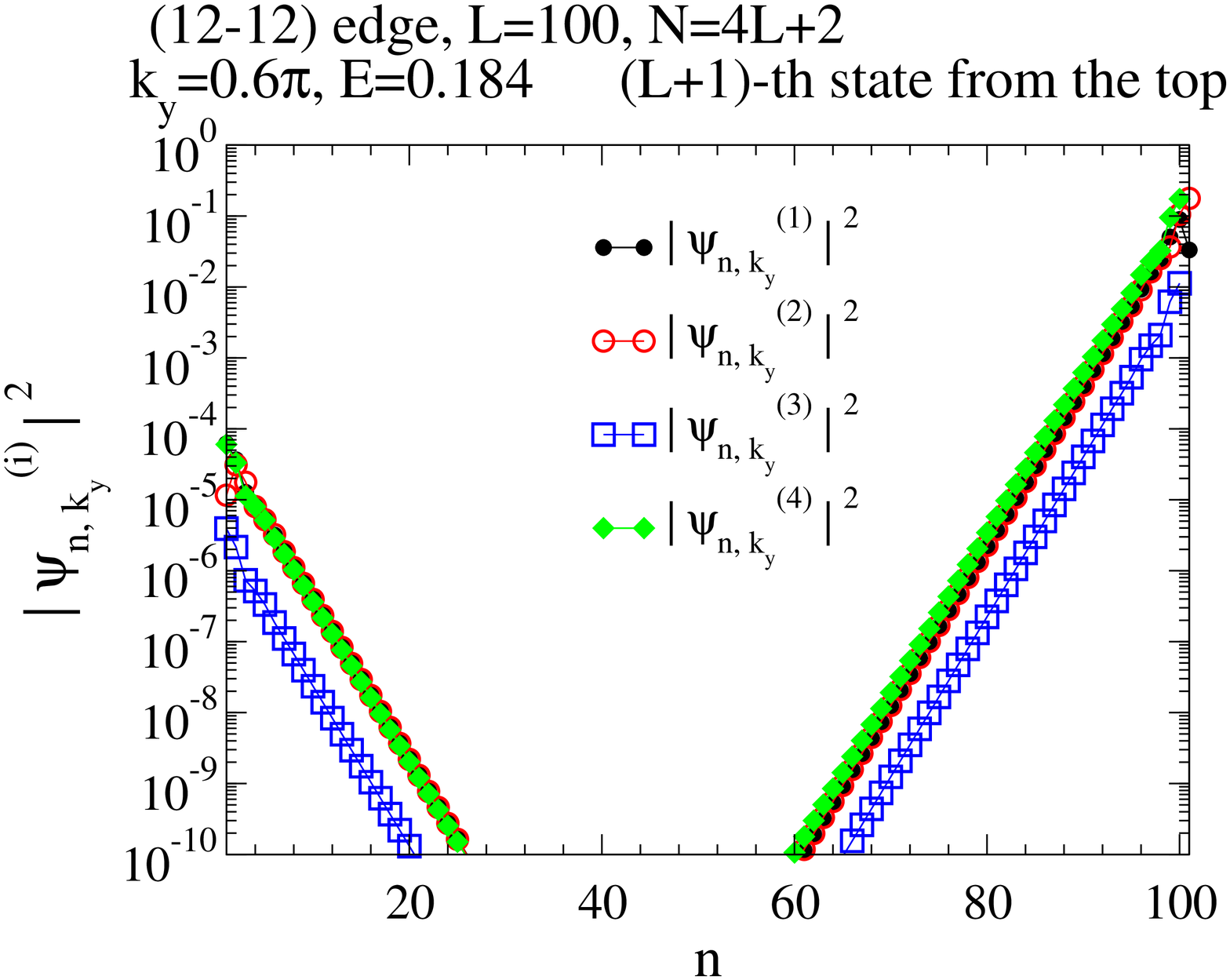}
\end{center}
\caption{(Color online)
Edge states near $\frac{3}{4}$ filling
for systems with $(12-12)$ edges with $L=100$.
The squares of the absolute value
$\Psi_{n,k_y}^{(i)}$ ($k_y=0.6\pi$) 
of the $L$-th (a) and $(L+1)$-th states from the top
are plotted as a function of $n$.
The values of $\Psi_{n,0.6\pi}^{(1)}$ 
$\Psi_{n,0.6\pi}^{(2)}$ and $\Psi_{n,0.6\pi}^{(4)}$ 
are almost the same in this scale of the figure.
The exponentially localized character is seen.}
\label{figwavef1212}
\end{figure}

There are $N=4 \times L$ eigenstates and energies for each $k_y$
for the (12-34) edge and the (34-12) edge.
We plot the energy as a function of $k_y$ in Fig.~\ref{figedge1234},
where we use different colors for each band of $L$ states.
There are edge states at the 1/4, 1/2 and 3/4 fillings of the band.
In this paper we focus on the edge state at 3/4 filling case,
since 3/4 filling is realized in $\alpha$-(BEDT-TTF)$_2$I$_3$.

In Fig.~\ref{figedgee0} we plot the energies near the $3/4$ filling
for the system with periodic boundary conditions for the x-direction,
the (12-34) edge, the (12-12) edge and the (34-34) edge.
If the system is periodic with respect to $x$ i.e., 
$\alpha=1$, we obtain the projection of the 3D plot of the energy
(Fig.~\ref{figfig3d}) as shown in Fig.~\ref{figedgee0}(a).

The eigenstates of the (12-34) edge at $k_y=0.6 \pi$ and $k_y=0$
for the $L$-th  state from the top and that at $k_y=0.6 \pi$
for the $(L+1)$-th  state
are shown in Fig.~\ref{figwavef1234} (a) , (b) and (c), respectively.
The $L$-th state from the top, i.e. 
the bottom of the highest band, is the edge state, which is
localized at the 
left edge for $|k_y|=0.6 \pi > |K_y| \approx 0.353 \pi$ 
(Fig.~\ref{figwavef1234} (a))
  and at the right edge 
for $|k_y| = 0 < |K_y|$ (Fig.~\ref{figwavef1234} (b)).
We define the localization length, $n_0$,  as 
\begin{equation}
|\Psi_{n,k_y}^{(i)}|^2 \propto \exp(-n/n_0),
\label{eqlength1}
\end{equation}
for the edge state localized at the left edge
and 
\begin{equation}
|\Psi_{n,k_y}^{(i)}|^2 \propto \exp(-(L-n)/n_0),
\label{eqlength2}
\end{equation}
for the edge state localized at the right edge.
It is obtained that
 $n_0 \approx 1.9$ at $k_y=0.6 \pi$ and 
$n_0 \approx 0.37$ at $k_y=0$, as seen 
in Fig.~\ref{figwavef1234}(a) and (b).
Other states such as 
the $(L+1)$-th state, for example, are not the edge state as seen in
 Fig.~\ref{figwavef1234}(c).

The edge states exist only at $|k_y| > |K_y|$ 
for the (12-12) edge and at  $|k_y| < |K_y|$ 
for the (34-34) edge, as seen in Fig.~\ref{figedgee0} (c) and (d).
In these cases the edge states are localized in both left and right
edges.
In general,
when the localization length, $n_0$,  of the edge state is
 much smaller than
the width $L$ ($n_0 \ll  L$),
the edge states at each edge can be treated as independent states,
and these states degenerate due to inversion symmetry. 
Otherwise, the edge states at each edge
interact each other and
the degeneracy is lifted. 
Then the ``bonding''
 and ``anti-bonding'' states of the each edge states
become the eigenstates. In our numerical calculation
for $L=100$, which is much larger than the localization length,
the $L$-th and the $(L+1)$-th states from the top have the same energy
within the numerical accuracy and the eigenstates are any linear combinations
of the left and the right eigenstates,
as shown in Fig.~\ref{figwavef1212}, where $L$-th and $(L+1)$-th states
are localized at both edges. 

From the above results 
we conclude that if the left or right edge
is the chain with sites 1 and 2, the edge states 
exists in the region $|k_y| > |K_y|$ near 3/4 filling.
  If the left or right edge
is the chain with sites 3 and 4, the edge states 
exists in the region $|k_y| < |K_y|$ near 3/4 filling.
We summarize the existence of the edge states
 in Table~\ref{table1} (a).
\begin{table}[b]
 \begin{center}
  \begin{tabular}{|cc|cc|}
   \multicolumn{3}{l}{(a) vertical edge}            \\ \hline
    left & edge state    & right & edge state   \\ \hline            
    12   & $|k_y| > |K_y|$ & 12    & $|k_y| > |K_y|$ \\ 
    34   & $|k_y| < |K_y|$ & 34    & $|k_y| < |K_y|$ \\ \hline
  \multicolumn{3}{c}{} \\
   \multicolumn{3}{l}{(b) horizontal edge}            \\ \hline
    lower & edge state    & upper & edge state   \\ \hline            
    14    & $|k_x| > |K_x|$ & 42    & $|k_x| > |K_x|$ \\ 
    31    & $|k_x| > |K_x|$ & 23    & $|k_x| > |K_x|$ \\ \hline 
    23    & $|k_x| < |K_x|$ & 31    & $|k_x| < |K_x|$ \\ 
    42    & $|k_x| < |K_x|$ & 14    & $|k_x| < |K_x|$ \\ \hline
  \end{tabular}
 \end{center}
\caption{
Existing region of the edge states
in $\alpha$-(BEDT-TTF)$_2$I$_3$
at 3/4 filling. 
See Fig.~\ref{figgraphene}(a).}
\label{table1}
\end{table}%
\begin{table}[b]
 \begin{center}
  \begin{tabular}{|cc|}
   \multicolumn{2}{l}    {vertical edge}        \\ \hline
    left, right & edge state      \\ \hline            
    zigzag     & $|k_y| > |K_y|$  \\ 
    bearded (Klein's edge\cite{Klein1994})    & $|k_y| < |K_y|$                   \\ \hline
  \multicolumn{2}{c}{} \\
   \multicolumn{2}{l}{horizontal edge}            \\ \hline
    lower, upper  & edge state      \\ \hline            
   armchair with Klein's edge\cite{Wakabayashi2010}   
& $|k_x| > |K_x|$ \\ 
    armchair   & $|k_x| < |K_x|$                 \\ 
\hline
  \end{tabular}
 \end{center}
\caption{Existing region of the edge states in graphene. 
For the isotropic honeycomb lattice, 
$|K_x|=0$ and $|K_y|=\frac{2}{3}\pi$.
Then no edge states exist
in the armchair edge.
 However, edge states in armchair edge exist
in the anisotropic case\cite{Kohmoto2007}, where $|K_x| \neq 0$.
See Fig.~\ref{figgraphene}(b) and (c).
}
\label{table2}
\end{table}%
\begin{figure}[bth]
\begin{center}
\vspace{5mm}
\includegraphics[width=0.38\textwidth]{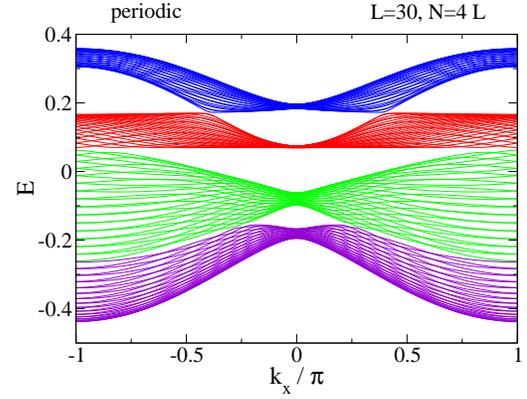}
\end{center}
\caption{(Color online)
Energy spectrum 
of $\alpha$-(BEDT-TTF)$_2$I$_3$ with periodic boundary conditions 
as a function of the wave number $k_x$.
The parameters are the same as in Fig.~\ref{figedge1234}.
}
\label{figedgeperiodicy}
\end{figure}
\begin{figure}[btp]
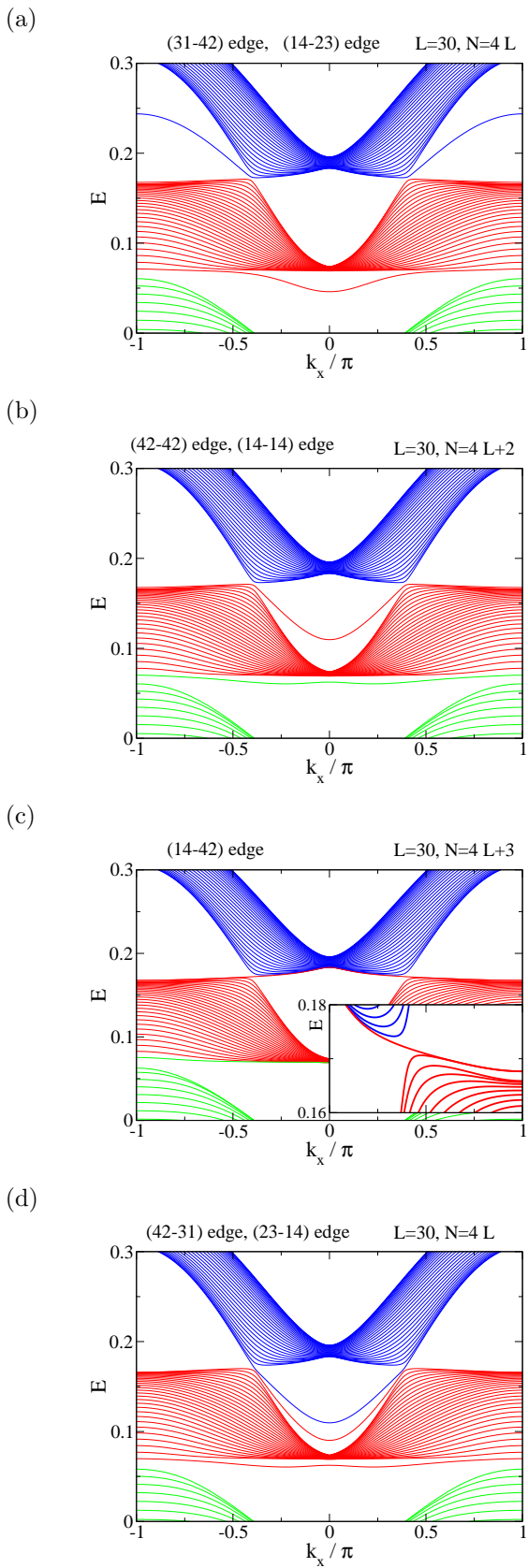

\begin{center}
\begin{flushleft}(a)\end{flushleft}
\includegraphics[width=0.35\textwidth]{64500fig8a.eps} 
\begin{flushleft}(b)\end{flushleft}
\includegraphics[width=0.35\textwidth]{64500fig8b.eps} 
\begin{flushleft}(c)\end{flushleft}
\includegraphics[width=0.35\textwidth]{64500fig8c.eps} 
\begin{flushleft}(d)\end{flushleft}
\includegraphics[width=0.35\textwidth]{64500fig8d.eps} 
\end{center}
\caption{(Color online) 
Energies near the 3/4 filling in some horizontal edges. 
The energies of the edge state at $|K| > |K_x|$
near the lower edge of sites 1 and 4 
or near the upper edge of sites 4 and 2 are close to the energies of the
top of the second band, as shown in (a), (b), (c) and the inset of (c),
which is the magnified figure.  }
\label{fighorizontal}
\end{figure}
\begin{figure}[btp]
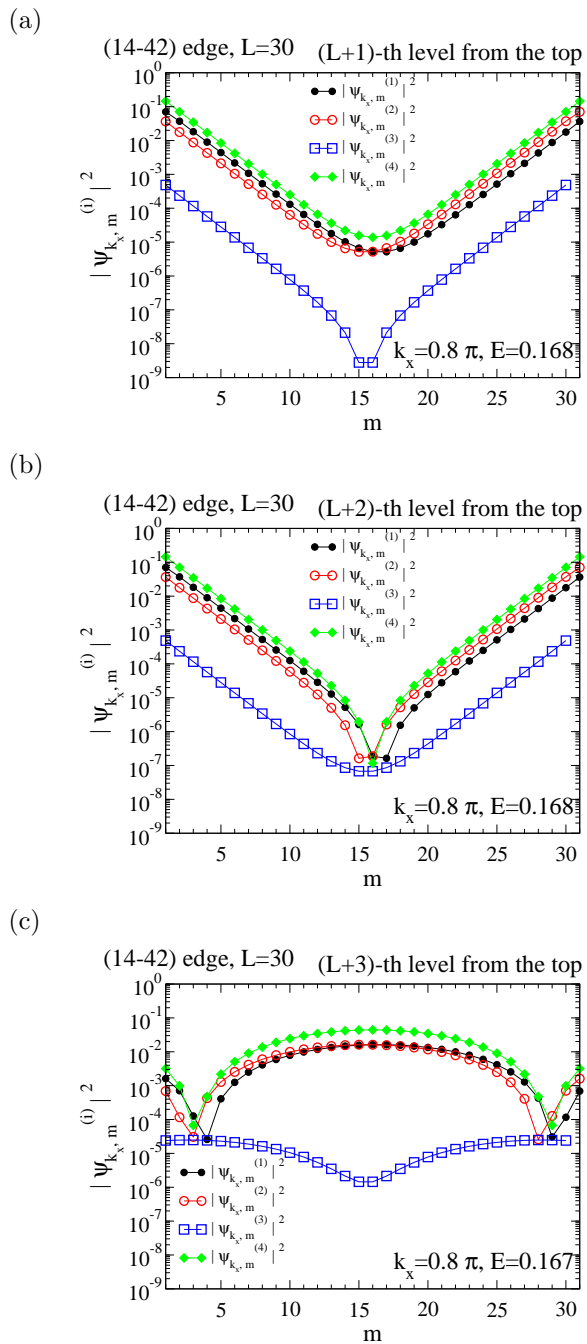

\begin{center}
\begin{flushleft}(a)\end{flushleft}
\includegraphics[width=0.38\textwidth]{64500fig9a.eps}
\\
\begin{flushleft}(b)\end{flushleft}
\includegraphics[width=0.38\textwidth]{64500fig9b.eps}
\\
\begin{flushleft}(c)\end{flushleft}
\includegraphics[width=0.38\textwidth]{64500fig9c.eps}
\end{center}
\caption{(Color online)
Wave function of the
(a) $(L+1)$-th, (b) $(L+2)$-th, and (c) $(L+3)$-th
states from the top of the energy for the (14-42) edge at $k_y=0.8 \pi$.
We take the same parameters parameters 
 as in Fig.~\ref{figedge1234}.
The $(L+1)$-th and $(L+2)$-th states are edge states, 
but $(L+3)$-th state is not
the edge state.
}
\label{figwave1442b}
\end{figure}

Next we study the  horizontal edges. 
In Fig.~\ref{figedgeperiodicy} 
we plot the energy of the
system with periodic boundary conditions as a function of $k_x$.
For various types of horizontal edges,  
we find that edge states always exist. 
 We plot some examples of energy of
the horizontal edges as a function of $k_x$ in Fig.~\ref{fighorizontal}. 

We obtain that in the 3/4-filled case
there are the edge states localized near the lower edge 
of the (14-xy) and (31-xy) edges when $|k_x| > |K_x|$,
where xy is either 14, 31, 23, or 42. 
Since  (14-xy) and (31-xy) edges are
changed to (y$'$x$'$-42) and (y$'$x$'$-23) edges 
by inversion, 
 where y$'$x$'$ is either 42, 23, 31, or 14, respectively,
(note that sites 1 and 2 are exchanged each other by inversion, 
while sites 3 and 4 are not changed.),
there are edge states localized near the upper edge
of (y$'$x$'$-42) and (y$'$x$'$-23) edges when $|k_x| > |K_x|$.

We also obtain that there are the edge states 
localized near the lower edge 
of the (23-xy) and (42-xy) edges when $|k_x| < |K_x|$,
and there are the edge states localized near the upper edge 
of the  (y$'$x$'$-31) and (y$'$x$'$-14) edges when $|k_x| < |K_x|$.
Table \ref{table1} (b) shows the summary of the 
existing region of the edge states for 
the horizontal edges.

The edge states in the same rows in table~\ref{table1}~(b), 
for example, the lower 14 
 and upper 42, have the same 
energy dispersion. 
The edge states in different rows in table~\ref{table1}~(b),
such as the lower 31 and upper 42,
 have different energies as a function of $k_x$.
In Fig.~\ref{fighorizontal}(a), we plot the energy as a function of the
$k_x$ in the (31-42) edge (the (14-23) edge has the same energy).
The $L$-th and the $(L+1)$-th
states  from the top of the energy  are edge states at $|k_x| > |K_x|$.
The edge states in the $(L+1)$-th state (at $|k_x| > |K_x|$)
 has a relatively
large localization length and the energy
of the edge state is close to the second band.
 The energy of the $(L+1)$-th state at $|k_x| > |K_x|$ 
in the (31-42) edge (Fig.~\ref{fighorizontal}~(a))
is same as the $(L+1)$-th state at $|k_x| > |K_x|$ in the (42-42) edge
(Fig.~\ref{fighorizontal}~(b)) and  
the $(L+1)$-the and the $(L+2)$-th states in the
(14-42) edge (Fig.~\ref{fighorizontal}~(c) and the inset).
In Fig.~\ref{figwave1442b} we plot the wave function
of the  $(L+1)$-the, the $(L+2)$-th and $(L+3)$-th states
in the (14-42) edge at $k_x=0.8 \pi$
as a function of the site number $m$.
It is clearly seen that the $(L+1)$-th state and the
$(L+2)$-th state are the edge states localized 
at both lower and upper edges,
while the $(L+2)$-th state is not the edge state.

The energies as a function of $k_x$ for the
 edge states of the (23-xy) edge and the (42-xy) edge are different,
as seen two curves between the first and
 the second bands ($L$-th and (L+1)-th states) from the top of the energy
at $|k_x| < |K_x|$ in Fig.~\ref{fighorizontal}~(d).
As summarized in Table~\ref{table1},
all edges have the edge states at $|k_y| > |K_y|$ or $|k_y| < |K_y|$
for the vertical edge and at $|k_x| > |K_x|$ or $|k_x| < |K_x|$
for the horizontal edge.

\begin{figure}[btp]
\begin{center}
\vspace{0.5cm}
\includegraphics[width=0.38\textwidth]{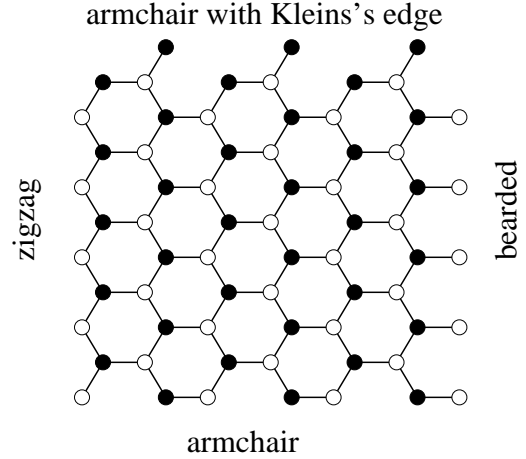}
\end{center}
\caption{(Color online)
Possible edges for the graphene (honeycomb lattice),
zigzag, bearded, armchair and armchair with the Klein's edge.
}
\label{figgrapheneedge}
\end{figure}

The edge states in $\alpha$-(BEDT-TTF)$_2$I$_3$ obtained above are
compared to the edge states in graphene with anisotropic hoppings.
In graphene there are two sites in the unit cell,
forming A and B sublattices.
These sites are exchanged by inversion. 
The edges in the graphene are classified as
zigzag, bearded, 
zigzag-bearded and
armchair\cite{Kohmoto2007}. We can also 
consider the armchair with the Klein's edge\cite{Wakabayashi2010}.
These edges are shown in Fig.~\ref{figgrapheneedge}.

\begin{figure}[btp]
\begin{center}
\begin{flushleft}
(a)
\end{flushleft}
\includegraphics[width=0.48\textwidth]{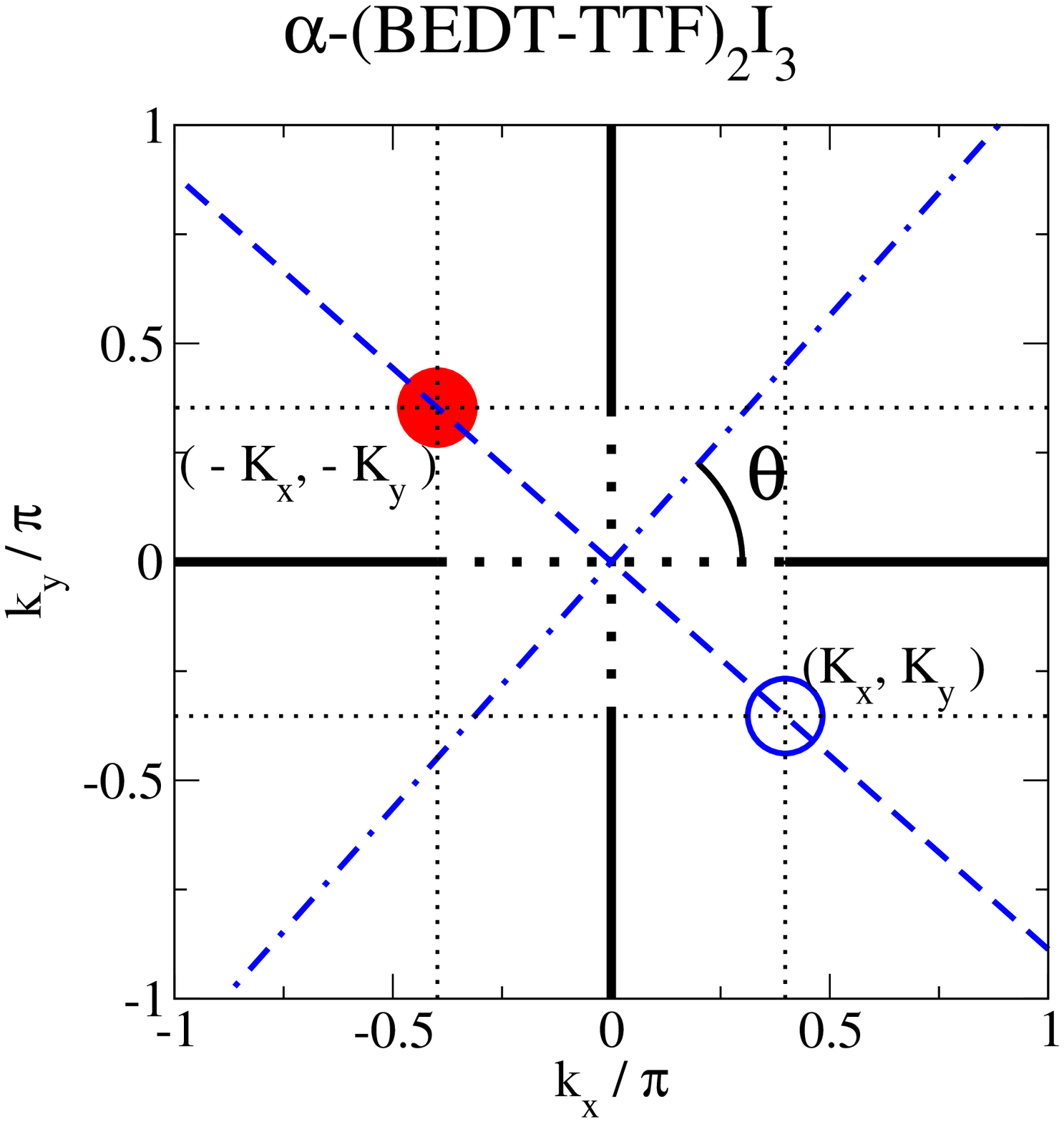}
\begin{flushleft}
(b)
\end{flushleft}
\includegraphics[width=0.48\textwidth]{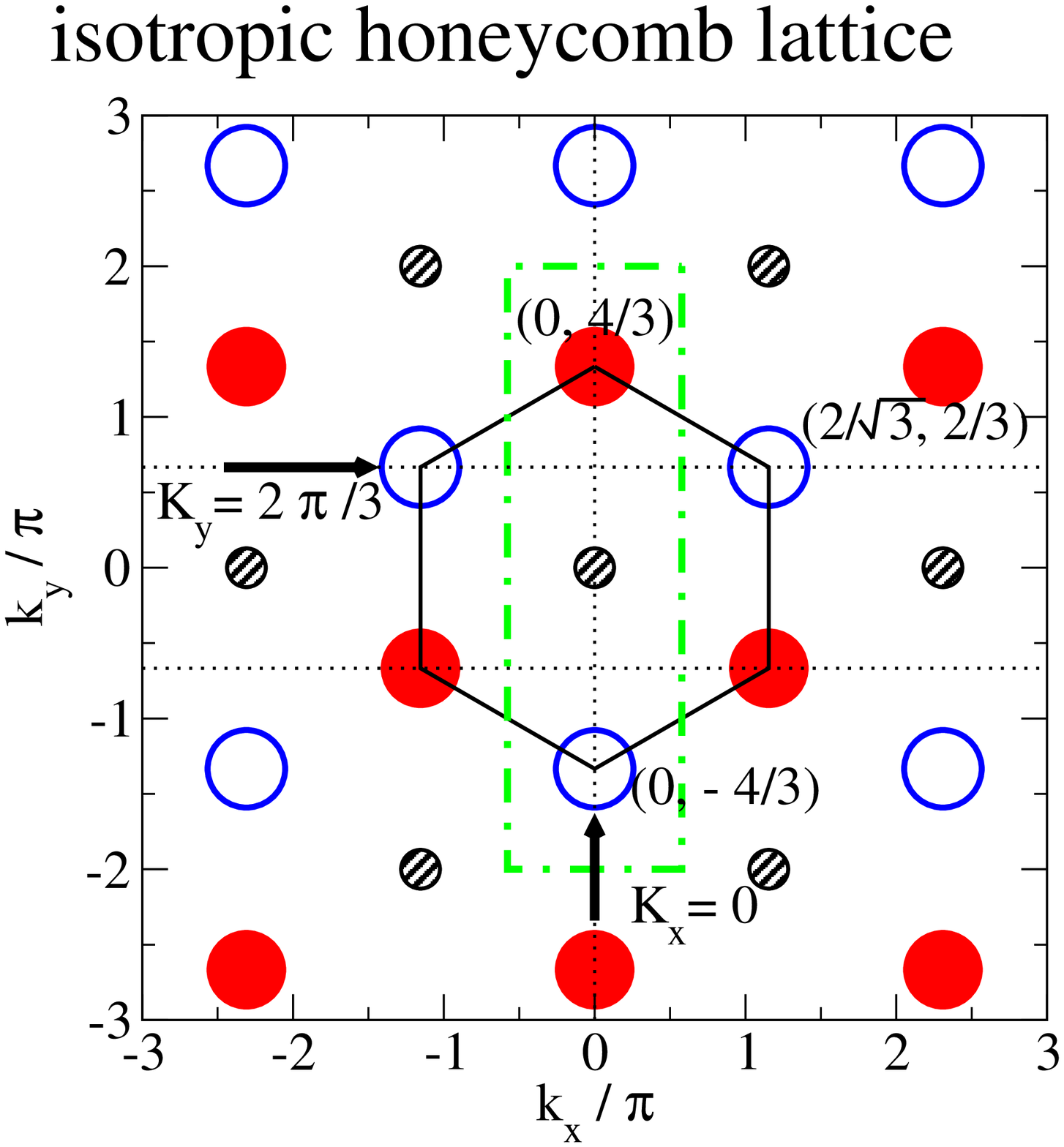}
\begin{flushleft}
(c)
\end{flushleft}
\includegraphics[width=0.48\textwidth]{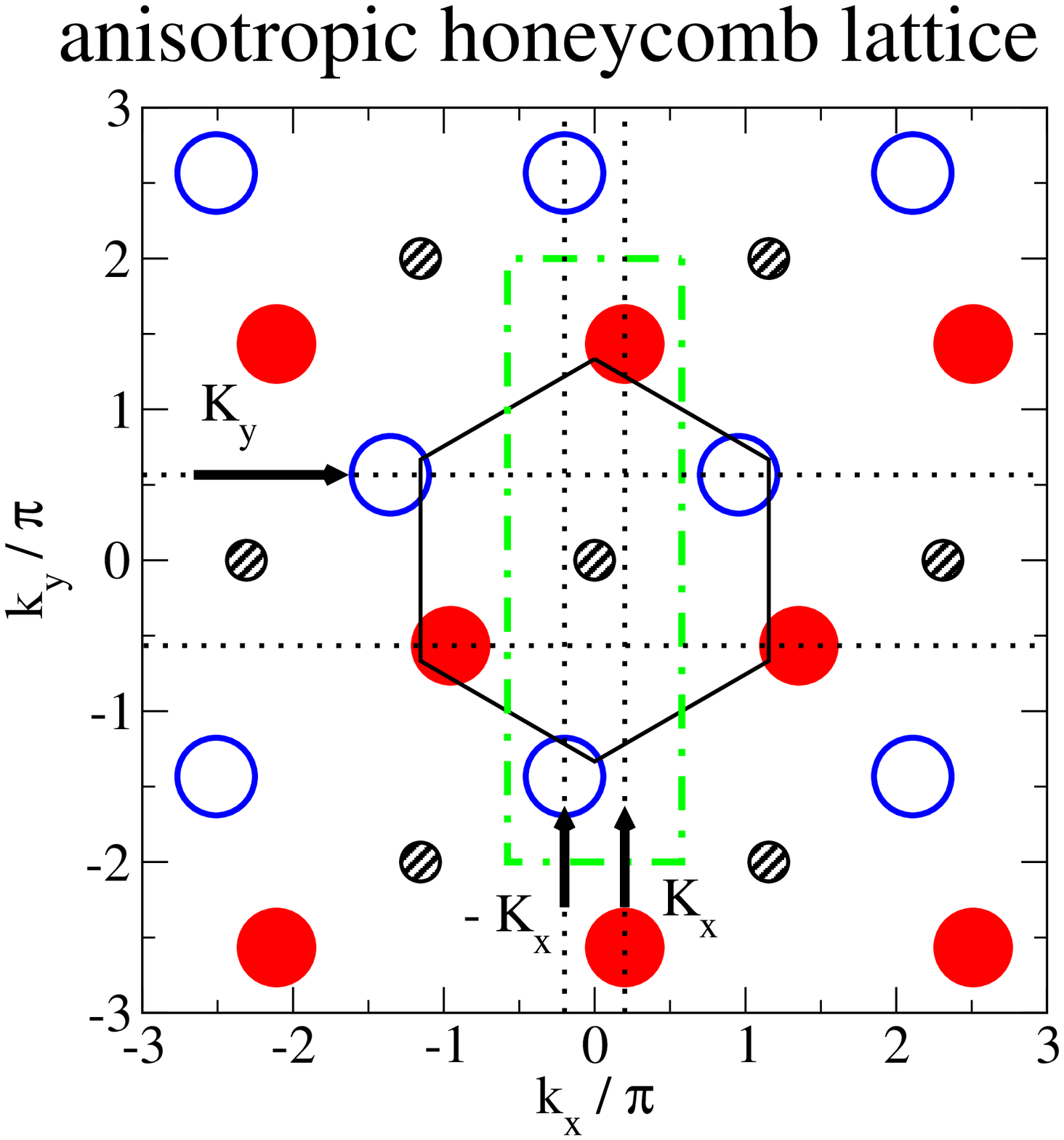}
\end{center}
\caption{(Color online)
The Dirac points 
for electrons on (a) $\alpha$-(BEDT-TTF)$_2$I$_3$, 
(b) isotropic honeycomb lattice, and (c) anisotropic  
honeycomb lattice. 
The small striped circles are 
$\Gamma$ points.
The Dirac points are shown by blue open circles and red filled circles. 
Hexagons in (b) and (c) are the Brillouin zone. 
The green rectangles in (b) and (c) have the same areas as hexagons. 
If the hoppings between the nearest sites are anisotropic,
the Dirac points move from the corners of the Brillouin zone
as shown in (b) and (c).
If the tilted edge in $\alpha$-(BEDT-TTF)$_2$I$_3$
can be made, the projections 
of the Dirac points into the direction 
of blue dot-dashed line in (a) coincide with each other.
}
\label{figgraphene}
\end{figure}

We consider the anisotropic cases with different hoppings between 
nearest neighbors in different directions\cite{Kohmoto2007}.
For the isotropic honeycomb lattice
the edge states exist in the zigzag, bearded and zigzag-bearded
 edges for $|k_y|>2\pi/3$, $|k_y|<2\pi/3$ and any $k_y$, 
respectively, while
 there is no edge states in the armchair edges.
The Dirac points in the isotropic honeycomb lattice 
are located at $(0, \pm 4\pi/(3 a) )$, where $a$ 
is a lattice constant of the honeycomb lattice, as shown 
in Fig.~\ref{figgraphene}(b).
If we consider the anisotropic cases, however,
it has been shown that 
the edge states exit even in the armchair edges\cite{Kohmoto2007}.
The reason for the 
absence of the edge states in the isotropic honeycomb lattice
with armchair edges can be understood as follows.
In the armchair edges, the edge states can exist in the region 
$|k_x| <|K_x|$
where $\pm K_x$ are the $x$ component of two Dirac points.
In the case of the isotropic honeycomb lattice, the projections of the two 
Dirac points into $k_x$ axis coincide at $K_x=0$. 
In that case 
there are no region of $k_x$, which satisfies $k_x < |K_x|=0$, 
and as a result there are no edge states in the isotropic armchair edges.
On the other hand, if we consider the anisotropic honeycomb lattice
 with $K_x \neq 0$, we have the edge states at $|k_x| < |K_x|$. 
Recently, it is also shown that
the edge states exist 
in the isotropic honeycomb lattice with armchair edges
moderated by the Klein's defects\cite{Wakabayashi2010}.
In Table~\ref{table2}, the exiting region of the edge states
are given. 

We speculate that if we can make the tilted edges
with the tilting angle $\theta=-\arctan (K_x/K_y)$, then
coincidence of the projections of two Dirac points 
in the wave number along the tilted direction,
as shown in Fig.~\ref{figgraphene}(a).
In that case we obtain the similar situation as in the isotropic
honeycomb lattice with armchair edges and we expect that the edge states 
cannot exist.
%

One of
the differences between the edge states in $\alpha$-(BEDT-TTF)$_2$I$_3$
and the edge states in graphene is that
all four components of the wave functions decays exponentially
away from the edge
in edge states in  $\alpha$-(BEDT-TTF)$_2$I$_3$, while 
one component of the wave function is always zero and the other
component decays exponentially in the edge states
in graphene. The other difference is the 
wave number dependence of the energy of the edge state.
The  energy of the edge states in 
$\alpha$-(BEDT-TTF)$_2$I$_3$ has strong $k_y$
or $k_x$ dependence, which
is in contrast with the edge states in graphene,
where the energy of the edge states is zero if the next-nearest-neighbor
hoppings are not taken into account.
The edge states in graphene have the $k_y$ dependent energy
only when the next-nearest-neighbor hoppings are 
finite.\cite{Sasaki2006,Peres2006} 

The energy dispersion of the edge states results in
the occupation of the edge states at $|k_y| < |K_y|$ 
in (12-34) edge when the system is 3/4 filled,
since the energy of the edge states is lower 
than the energy at the Dirac points,
as seen in Fig.~\ref{figedgee0} (b). 
In that case only the edge states at the right edge are occupied.
Similarly, the energy of the edge states for the horizontal edge have 
a wave-number dependence, as seen in Fig.~\ref{fighorizontal}.

\section{Site energy}
\begin{figure}[btp]
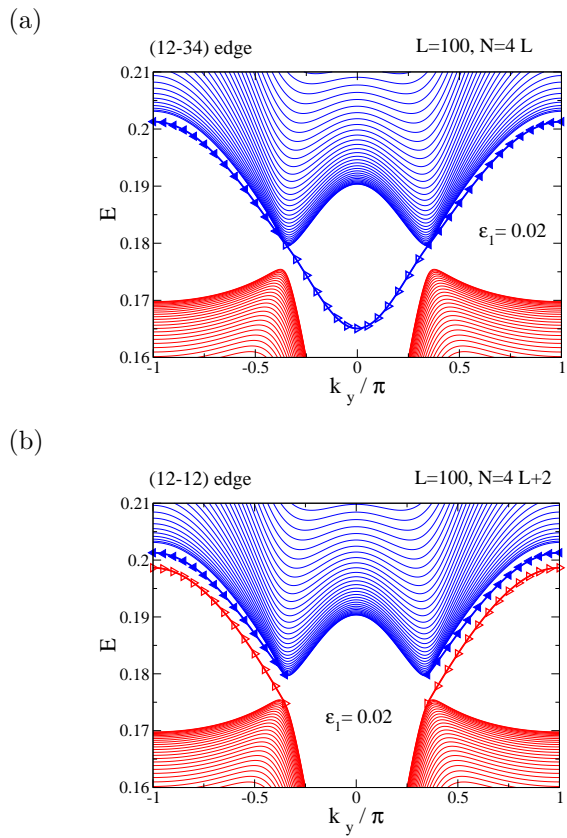

\begin{center}
\begin{flushleft}(a)\end{flushleft} 
\includegraphics[width=0.35\textwidth]{64500fig12a.eps}
\begin{flushleft}(b)\end{flushleft} 
\includegraphics[width=0.35\textwidth]{64500fig12b.eps}
\end{center}
\caption{(Color online)
Energy spectrum near $\frac{3}{4}$ filling
for systems with (12-34) and (12-12) edges.
The parameters are the same as in Fig.~\ref{figedge1234}
except for $\epsilon_1=0.02$,
which violate the inversion symmetry.
}
\label{figedgee102}
\end{figure}
\begin{figure}[btp]
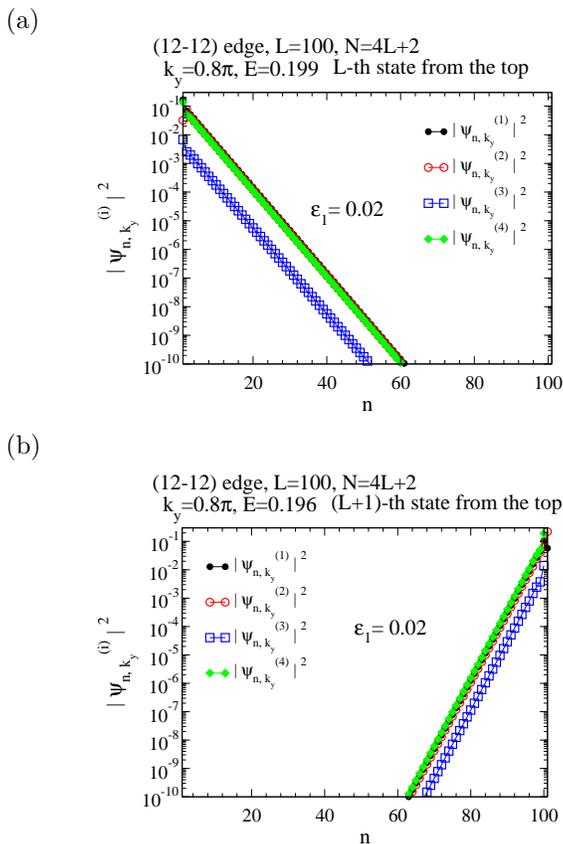

\begin{center}
\begin{flushleft}(a)\end{flushleft} 
\includegraphics[width=0.35\textwidth]{64500fig13a.eps}
\begin{flushleft}(b)\end{flushleft} 
\includegraphics[width=0.35\textwidth]{64500fig13b.eps}
\end{center}
\caption{(Color online)
Wave functions of the $L$-th and $(L+1)$-th states from the top
at $k_y=0.8 \pi$
for systems with (12-12) edge.
The parameters are the same as in Fig.~\ref{figedge1234}
except for $\epsilon_1=0.02$,
which violate the inversion symmetry.
The $L$-th state from the top is localized at the left edge
with the localization length $n_0 \approx 2.90$ and 
the $(L+1)$-th state from the top is localized at the right edge 
with the localization length $n_0 \approx 1.80$.
}
\label{figwavee102}
\end{figure}
\begin{figure}[btp]
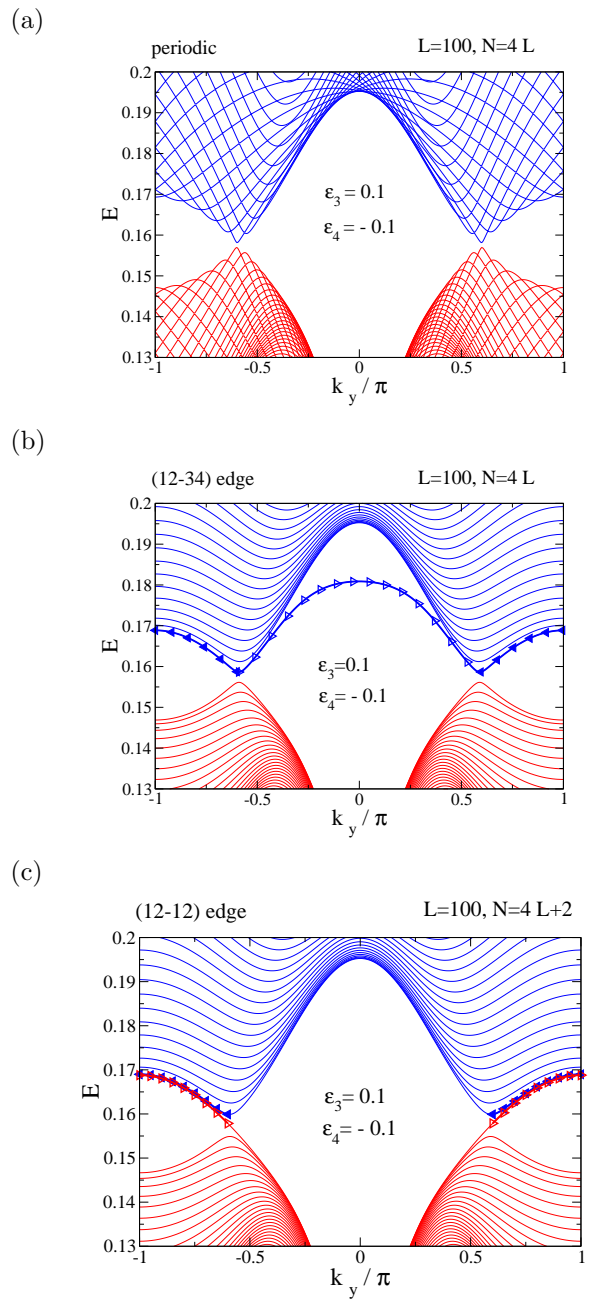

\begin{center}
\begin{flushleft}(a)\end{flushleft}
\includegraphics[width=0.35\textwidth]{64500fig14a.eps}
\begin{flushleft}(b)\end{flushleft}
\includegraphics[width=0.35\textwidth]{64500fig14b.eps}
\begin{flushleft}(c)\end{flushleft}
\includegraphics[width=0.38\textwidth]{64500fig14c.eps}
\end{center}
\caption{(Color online)
Energy spectrum near $\frac{3}{4}$ filling
for systems with  periodic boundary and with edges.
The parameters are $\epsilon_3=0.1$ and $\epsilon_4=-0.1$.}
\label{fig9abcd}
\end{figure}
\begin{figure}[btp]
\begin{center}
\includegraphics[width=0.4\textwidth]{64500fig15.eps}
\end{center}
\caption{(Color online)
Edge states near $\frac{3}{4}$ filling
for systems with $(12-34)$ edges with $L=100$.
The squares of the absolute value
$\Psi_{n,k_y}^{(i)}$ ($k_y=0$) 
of the $L$-th  state from the top
is plotted as a function of $n$.
It is  seen that this state is an edge state localized 
at the right edge with small localization length
$n_0 \approx 0.83$.}
\label{fig10}
\end{figure}
\begin{figure}[btp]
\begin{center}
\includegraphics[width=0.4\textwidth]{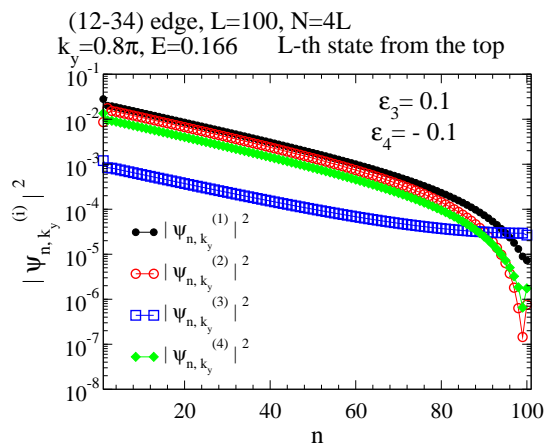}
\end{center}
\caption{(Color online)
Edge states near $\frac{3}{4}$ filling
for systems with $(12-34)$ edges with $L=100$.
The squares of the absolute value
$\Psi_{n,k_y}^{(i)}$ ($k_y=0$) 
of the $L$-th  state from the top
is plotted as a function of $n$.
It is seen that this state is an edge state localized 
at the left edge with the localization length $n_0 \approx 20.4$, $20.3$,
$21.5$, and $20.3$ for $|\Psi_{n,k_y}^{(i)}|$ with $i=1 - 4$, respectively,,
which are obtained from the least square fit of $\log |\Psi_{n,k_y}^{(i)}|$
for $5\leq n \leq 20$.}
\label{fig11}
\end{figure}
\begin{figure}[btp]
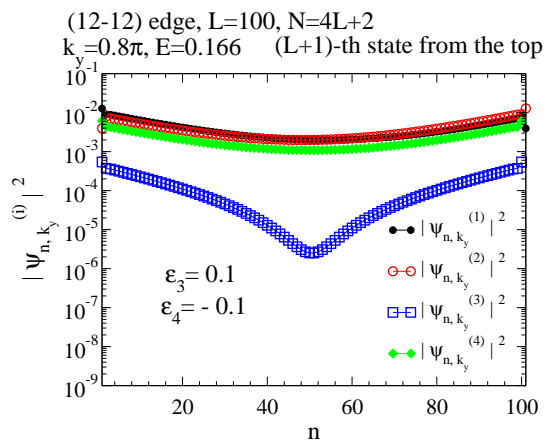

\begin{center}
\begin{flushleft}(a)\end{flushleft}
\includegraphics[width=0.4\textwidth]{64500fig17a.eps}
\begin{flushleft}(b)\end{flushleft}
\includegraphics[width=0.4\textwidth]{64500fig17b.eps}
\end{center}
\caption{(Color online)
Edge states near $\frac{3}{4}$ filling
for systems with $(12-12)$ edges with $L=100$.
The squares of the absolute value
$\Psi_{n,k_y}^{(i)}$ ($k_y=0.6\pi$) 
of the $L$-th (a) and $(L+1)$-th (b) states from the top
are plotted as a function of $n$.
The exponentially localized character is seen.
The localization lengths are estimated as
$n_0 \approx 17.4$, $15.8$, $50.0$ and $16.4$ for 
 $\Psi_{n,k_y}^{(i)}$ of the $L$-th state from the top with
$i =1 - 4$, respectively, and
 $n_0 \approx 22.8$, $24.4$, $14.2$ and $23.7$ for these of the
$(L+1)$-th state.
}
\label{fig12}
\end{figure}
In this section we study the effect of the site energies $\epsilon_i$.
Note that $\epsilon_3$ and $\epsilon_4$ do not violate the inversion
symmetry but $\epsilon_1$ and $\epsilon_2$ violate the inversion symmetry
if $\epsilon_1 \neq \epsilon_2$. 
When $\epsilon_1=\epsilon_2=\epsilon_3=\epsilon_4$,
the dispersion of the eigenstates 
is only shifted by the constant energy.
Therefore, if 
we consider the system with inversion symmetry, 
$\epsilon_3$ and $\epsilon_4$ are the relevant parameters. Either $\epsilon_1$
or $\epsilon_2$ can be taken as a 
parameter for the breaking of the inversion symmetry.

The energy gap is opened 
when inversion symmetry is broken by $\epsilon_1$ or $\epsilon_2$, 
as shown in Fig.~\ref{figedgee102}.
Even if the gap is opened by  $\epsilon_1$ or $\epsilon_2$,
 the edge states still exist.
As seen in Fig.~\ref{figedgee102} (a) the $L$-th state from the top
in the system with (12-34) edge
is the left edge state for $|k_y| > K_y$ and the right edge state for 
$|k_y| < K_y$, which is the same as in the system with inversion symmetry
(see Fig.~\ref{figedgee0}(b)). The (12-12) edge has the edge states
for  $|k_y| > K_y$ as shown in Fig.~\ref{figedgee102} (b).
Because the inversion symmetry is broken, the left and right edge states are 
not degenerated. The $L$-th state from the top of the energy
is the edge state localized at the left edge, and the
$(L+1)$-th state is the edge state at the right edge.
The localization length of these edge states are different
as shown in Fig.~\ref{figwavee102} (a) and (b).

As seen in Fig.~\ref{figedgee102}, the edge states with $\epsilon_1 \neq 0$
have the similar strong $k_y$ dependent energy as in the system 
with inversion symmetry. As a result, if the system is 3/4 filled, i.e.
the chemical potential is in the middle of the gap
of the bulk system (for example $\mu=0.175$ in Fig.~\ref{figedgee102} 
(a) and (b)), the localized state at the right edge is partially filled.

 The first and the second bands from the top of the energy
 still touch each other at the Dirac points
even when we take finite $\epsilon_3$ and $\epsilon_4$
(as an example we take $\epsilon_3=0.1$ and $\epsilon_4=-0.1$), 
as shown in Fig.~\ref{fig9abcd} (a) for the periodic boundary conditions.
In that parameters ($\epsilon_3=0.1$ and $\epsilon_4=-0.1$),
the edge states exist as in the case of $\epsilon_3=\epsilon_4=0$
(see Fig.~\ref{figedgee0} and Fig.~\ref{fig9abcd}).
The localization length, however, differs  from that in the case of
$\epsilon_3=\epsilon_4=0$. 
While the $L$-th state from the top 
at $k_y=0$ in the (12-34) edge is the edge state localized at the right edge 
with the localization length $n_0 \approx 0.83$ (Fig.~\ref{fig10}),  
the  $L$-th state at $k_y=0.8 \pi$ is the edge state at
the left edge with the large localization length $n_0 \approx 20$
(Fig~\ref{fig11}). 
The large localization length is more clearly seen 
in the (12-12) edge. The $L$-th and $(L+1)$-th states at $k_y=0.8 \pi$
are almost degenerate in energy in the system with $L=100$.
However the wave function of the $L$-th state are different 
from that of the $(L+1)$-th state   
as shown in Fig.~\ref{fig12}. 
The square of the absolute values of 
the wave functions, $|\Psi_{n,k_y}^{(i)}|^2$,
of $L$-th and $(L+1)$-th states from the top at $k_y=0.8 \pi$
are asymmetric and symmetric with respect to $n$
and they look like ``anti-bonding'' and ``bonding'' states,
respectively.
This means that these states are 
indeed the localized edge states but
the system size  ($L=100$) is not large enough for
the edge states to be treated as independent states at each edges.
In these cases the estimation of the localization length 
by Eqs.~(\ref{eqlength1}) or (\ref{eqlength2})
may have an ambiguity and should be taken carefully.
The localization lengths for $|\Psi_{n,k_y}^{(i)}|^2$ should be same for
all $i$ but 
depend on the component $i$
as we give the estimation in the figure captions.
Even in these cases, the existing regions of the edge states are
given in Table~\ref{table1} with $K_y$ varied
according to the choice of $\epsilon_3$ and $\epsilon_4$.

\clearpage
\section{Summary and discussions}
In this paper we study the edge states 
in $\alpha$-(BEDT-TTF)$_2$I$_3$
in the absence of the magnetic field, theoretically.
We have shown that the edge states exist in the
3/4 filled band in the system with four sites in the unit cell, 
such as $\alpha$-(BEDT-TTF)$_2$I$_3$.
We study the vertical and horizontal edges.
We show that all edges have the edge states as summarized in
Table~\ref{table1}.
In the edge states all components of the wave function
decays exponentially as a function of the distance from the edge.
Since the edge state has the
wave-number dependent energy, it is possible to
occupy only the left edge state or the lower edge states
at 3/4 filling system. For example,
in the (12-34) or (34-12) edges
only the localized state at the right edge or the left edge
is occupied in 3/4 filling system, respectively.

We also study the effect of the site energies, $\epsilon_1$, $\epsilon_2$, 
$\epsilon_3$ and $\epsilon_4$. When the inversion symmetry is broken by
$\epsilon_1 \neq 0$ or $\epsilon_2 \neq 0$, a finite gap appears
at the Dirac points. Even in that case the edge states
exist as in the system with inversion symmetry.
Due to the $k_y$ dependence of the edge states, the energy of the 
edge states
can locate in the middle of the bulk band gap.
Therefore, the insulator at the bulk sample has the partially 
filled edge states.
These states are not the topologically 
protected edge states, which are
recently predicted\cite{Kane2005,Bernevig2006,Fu2007} 
and observed\cite{Konig2007,Hsieh2008,Konig2008} 
in the topological insulators.

The opening of the gap at the Dirac points can be possible
either by the breaking of the inversion symmetry 
(in the honeycomb lattice with next-nearest-neighbor hoppings
the necessary symmetry for the zero gap is not the inversion
symmetry but the 
``averaged inversion symmetry''.\cite{Kishigi2008,Kishigi2008b})
or by the merging of 
the Dirac points.\cite{Hasegawa2006,Kobayashi2007,Montambaux2009}
If the inversion symmetry is not broken in (BEDT-TTF)$_2$I$_3$,
the opening of the gap at the Dirac points occurs only when
two Dirac points merge at $\mathbf{k}=-\mathbf{k}$ 
modulo a reciprocal vector,
i.e. $(k_x,k_y)=(0,0)$, $(0,\pi)$, $(\pi,0)$ or $(\pi,\pi)$,
which is predicted to occur at high pressure\cite{Kobayashi2007}.
In the present choice of parameters, the finite gap at 
the half-filling as seen in Fig.~\ref{figfig3d}
is caused by the merging of the Dirac points at $(\pi, 0)$.
The edge states also exist even when the gap is opened either by 
the breaking of the inversion symmetry or the merging of the Dirac points.

With finite $\epsilon_3$ 
and $\epsilon_4$ the gap remains zero at the Dirac points, but
the localization lengths of the edge states may be changed drastically.
The edge states with the large localization length can be realized
if we can tune the site energies.  

The experimental observation of the edge states 
in $\alpha$-(BEDT-TTF)$_2$I$_3$ will be possible
and provide us more insight about the massless Dirac particles
realized in the quasi-two-dimensional conductors.

%
\end{document}